\documentclass[11pt,a4paper, twocolumn]{aa}
\usepackage[utf8]{inputenc}
\usepackage[T1]{fontenc}
\usepackage{lmodern}
\usepackage{microtype}
\usepackage{setspace}
\usepackage{graphicx}
\usepackage{booktabs}
\usepackage{hyperref}
\hypersetup{colorlinks=true,linkcolor=blue,citecolor=blue,urlcolor=blue}
\usepackage{titling}
\usepackage{fancyhdr}
\usepackage{amsmath,amssymb}
\usepackage{multicol}

\usepackage{xcolor}
\usepackage{soul}

\begin{document}

\begin{titlepage}
  \centering
  \vspace*{1cm}
  {\Huge\bfseries Expanding Horizons \\[6pt] \Large Transforming Astronomy in the 2040s \par}
  \vspace{1.5cm}

    {\LARGE \textbf{The Galactic White Dwarf Population}\par}
  \vspace{1cm}

  \begin{tabular}{p{4.5cm}p{10cm}}
    \textbf{Scientific Categories:} & (white dwarfs: populations, evolution) \\
    \\
    \textbf{Submitting Author:} & Name: Santiago Torres \\
    & Affiliation: Departament de Física, Universitat Politècnica de Catalunya, c/Esteve Terrades 5, 08860, Castelldefels, Spain  \\
    & Institut d'Estudis Espacials de Catalunya (IEEC), C/Esteve Terradas, 1, Edifici RDIT, 08860, Castelldefels, Spain \\
    & santiago.torres@upc.edu\\

    \\
    \textbf{Contributing authors:} & Roberto Raddi$^{1}$, Alberto Rebassa-Mansergas$^{1,2}$, Leandro G. Althaus$^1$, Maria Camisassa$^1$, Tim Cunningham$^3$, Camila Damia Rinc\'on$^1$, Aina Ferrer i Burjachs$^1$, Nicola Gentile Fusillo$^4$, Enrique Garc\'ia-Zamora$^1$, Anna F. Pala$^5$, Steven Parsons$^6$, Ingrid Pelisoli$^7$, Nicole Reindl$^8$,
    Snehalata Sahu$^7$, Alejandro Santos-Garc\'ia$^1$, Pier-Emmanuel Tremblay$^7$, Odette Toloza$^{9,10}$\\
\\
\multicolumn{2}{l}{\small $^{1}$Departament de Física, Universitat Politècnica de Catalunya, c/Esteve Terrades 5, 08860, Castelldefels, Spain}\\
\multicolumn{2}{l}{\small $^{2}$Institut d'Estudis Espacials de Catalunya (IEEC), C/Esteve Terradas, 1, Edifici RDIT, 08860, Castelldefels, Spain}\\
\multicolumn{2}{l}{\small $^{3}$Center for Astrophysics, Harvard \& Smithsonian, 60 Garden St., Cambridge, MA 02138, USA} \\
\multicolumn{2}{l}{\small $^{4}$Universita degi studi di Trieste, Via Valerio, 2, Trieste, 34127, Italy} \\
\multicolumn{2}{l}{\small $^{5}$European Southern Observatory, Karl Schwarzschild Straße 2, D-85748, Garching, Germany}\\
\multicolumn{2}{l}{\small $^{6}$strophysics Research Cluster, School of Mathematical and Physical Sciences, University of Sheffield, Sheffield S3 7RH, UK}\\
\multicolumn{2}{l}{\small $^{7}$Department of Physics, University of Warwick, Coventry, CV4 7AL, UK}\\
\multicolumn{2}{l}{\small $^{8}$Landessternwarte Heidelberg, Zentrum für Astronomie, Ruprecht-Karls-Universität, Königstuhl 12, 69117, Heidelberg, Germany}\\
\multicolumn{2}{l}{\small $^{9}$Departamento de Física, Universidad Técnica Federico Santa María, Avenida España 1680, Valparaíso, Chile}\\
\multicolumn{2}{l}{\small $^{10}$Millennium Nucleus for Planet Formation, NPF, Valparaíso, 2340000, Chile}\\  
  \end{tabular}

  \vspace{1cm}

  \vspace{0.5em}
  \begin{minipage}{0.9\textwidth}
    \small
    
  \textbf{Abstract:} 
The ESA {\em Gaia} mission has revolutionized our understanding of the white dwarf population, delivering an unprecedented census of these nearby remnants and revealing previously unseen structures in the Hertzsprung–Russell (HR) diagram. However, while {\em Gaia} has expanded the scope of white dwarf astrophysics, it has also exposed new questions related to atmospheric composition, spectral evolution, crystallization, magnetism, and merger-driven pathways. Many of these open problems are encoded in the detailed morphology of the {\em Gaia} HR diagram, where precise spectroscopic characterization is essential for interpreting the underlying physical processes. Spectroscopic characterization, obtainable with current and future ESO facilities, can provide the effective temperatures and surface gravities that are required to derive accurate white dwarf masses, cooling ages, and luminosities. These fundamental parameters not only enable studies of spectral evolution, interior physics, and the origin of magnetic and high-mass white dwarfs, but also guarantee the construction of robust mass distributions and luminosity functions, essential for constraining the initial-to-final mass relation, probing the initial mass function, and reconstructing the star formation history of the local Galaxy, among other applications. Looking toward the 2040s, future multi-fiber spectrographs operating in survey mode on 10--15 meter class telescopes will be able to collect a complete spectroscopic sample of white dwarf, enabling the detailed characterization of their population. Achieving spectroscopic completeness for the nearby Galactic population and securing high signal-to-noise, moderate-to-high resolution spectra across the HR diagram with ESO instrumentation will be critical steps toward resolving these longstanding questions in white dwarf astrophysics.
  
  \end{minipage}

\end{titlepage}


\section{Introduction and Background}
\label{sec:intro}

White dwarfs are the most common stellar remnants, originating from the burnt out ashes of low- to intermediate-mass stars \citep[$\leq 10$\,M$_\odot$;][]{doherty2017,cummings2018}. 
The thermal evolution of white dwarfs is governed by the cooling of their electron-degenerate He, CO, or ONe cores \citep{althaus2010}, with additional energy release due to internal processes such as crystallization, phase separation, element sedimentation, and chemical distillation and exsolution \citep{bauer2020,blouin2021,bedard2024b,camisassa2024}. Their masses and atmospheric compositions are set by the prior evolution of the progenitor star and by processes in the outer layers such as diffusion, mixing, and accretion \citep[e.g.,][and references therein]{bedard2024}.

Historically difficult to identify due to their Earth-like radii and, thus, low luminosities, white dwarfs have surged to a more prominent role thanks to the photometric depth and astrometric precision of the ESA {\em Gaia} mission \citep{gaia2016}. Within its latest data release, about 1.3 million candidates have been found with around $\simeq360,000$ being high-confidence white dwarfs \citep{gentilefusillo2021}. This order of magnitude increase with respect to the previous decade has led to many discoveries, which confirm earlier predictions, but also led to new challenges \citep[e.g.,][for a recent review]{tremblay2024}. 

Although most of the white dwarfs detectable by {\em Gaia} can be found out to $\sim500\,$pc, the 100-pc sample remains the most nearly volume-complete representation of the local population \citep[e.g.][]{jimenezesteban2023}. The spectroscopic census is almost complete within 40 pc \citep{tremblay2020,mccleary2020,obrien2023,obrien2024}, but at larger distances many objects still lack spectroscopic confirmation, despite recent progress from automatic classification methodologies \citep{vincent2024,garcia-zamora2025}.

\begin{figure}
    \centering
    \includegraphics[width=\linewidth]{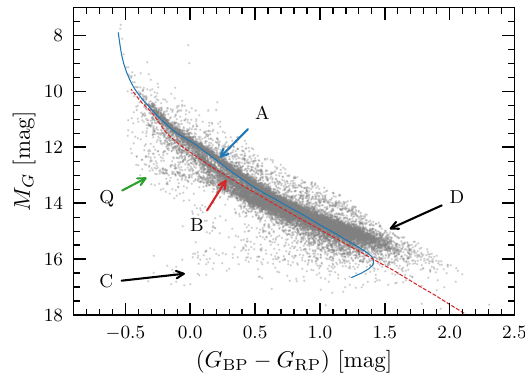}
    \caption{The {\em Gaia} white dwarf sample within 100-pc from the Sun. The cooling sequence of 0.6\,M$_\odot$  pure-H and 0.7\,M$_\odot$  pure-He atmospheres are shown by blue solid and red dashed lines, respectively \citep{camisassa2016,camisassa2017}. The locations of the A, B, and Q branches, as well as the faint ``blue" branch (C) and the ``red-excess'' (D) are also labeled.}
    \label{fig:hr}
\end{figure}

In Fig.\,\ref{fig:hr}, a clean, almost-complete, volume-limited sample within 100 pc from the Sun is shown in the {\em Gaia} Hertzsprung–Russell (HR) diagram \citep{gentilefusillo2021,jimenezesteban2023}. This diagram reveals the main features of the white dwarf cooling sequence, reflecting the fundamental physics and evolutionary processes of white dwarfs:

\begin{itemize}
\item[i)] the A branch, dominated by objects with H-rich atmospheres and coinciding with the cooling sequence of canonical $\approx 0.6$\,M$_\odot$ white dwarfs;
\item[ii)] the B branch, fainter than the A branch and overlapping the region of H-rich $\approx 0.7$--$0.8$\,M$_\odot$ white dwarfs, but predominantly populated by He-rich and mixed H/He atmospheres, a feature to which stealth carbon enrichment is believed to contribute significantly \citep{camisassa2023}; 
\item[iii)] the Q branch, a nearly horizontal structure of the {\em Gaia} HR diagram associated with the crystallization pile-up, showing a relative higher proportion of carbon-atmosphere white dwarfs, high-mass merger products, and strongly magnetic objects \citep{tremblay2019,cheng2019,camisassa2021}.
\end{itemize}

In addition to these three main branches, the {\em Gaia} HR diagram also shows other puzzling features: a faint “blue” sequence at the coolest temperatures, generally attributed to the infrared-flux suppression caused by enhanced collision-induced absorption in helium-dominated atmospheres \citep{bergeron2022}, and a distinct red-excess locus at the faint end of the cooling track, which may arise from deficiencies in atmospheric opacity calculations for cool white dwarfs \citep{obrien2024,sahu2025}.

\section{Key Science Drivers in the 2040s}
\label{sec:openquestions}

The diverse features revealed by the {\em Gaia} HR diagram --including the A, B, and Q branches, as well as the infrared-deficit and red-excess sequences---can ultimately be understood in terms of several fundamental distributions of the white dwarf population, such as the spectral-type\footnote{White dwarf spectral types are grouped into DAs, which exhibit hydrogen absorption features, and non-DAs, which lack detectable hydrogen lines. Among the non-DAs, DBs show helium features, DQs carbon, DZs metals, and DCs no spectral features at all. Mixed or transitional types also exist, e.g., DBA stars with helium-dominated spectra and weak hydrogen lines, among other. See \citet{mccook1999} for more details.} distribution, the mass distribution, the luminosity function, and the magnetic field distribution. Kinematic information adds a complementary dimension that, together with spectral types and key physical parameters---effective temperature, surface gravity, mass, luminosity, and magnetic field--provides a coherent framework for understanding the Galactic components and evolutionary channels of the white dwarf population. In what follows, we briefly describe the most relevant of these distributions.

\subsection{The Spectral-Type Distribution}

A spectroscopically complete white dwarf sample will enable the first unbiased determination of the spectral-type distribution in the solar neighborhood. This distribution will encode the relative proportions of DAs, DBs, DCs, DQs, DZs, and mixed types, thereby reflecting the complex processes governing spectral evolution, including convective mixing, dilution, accretion of H and planetary debris, magnetic field emergence, and chemical diffusion. Understanding how these spectral fractions vary with effective temperature, mass, and kinematics  will clarify the evolutionary channels that link different atmospheric compositions \citep[e.g.][]{torres2023, bedard2024}.

\subsection{The Mass Distribution}

Accurate spectro-photometric determinations of effective temperature and surface gravity, combined with evolutionary models, yield precise white dwarf masses \citep{tremblay2019b,bergeron2019}. A precise and unbiased determination of the mass distribution has been a longstanding problem in white dwarf astrophysics, given that this distribution provides one of the most powerful diagnostics of the endpoints of stellar evolution. A spectroscopically complete sample will allow robust identification of the main characteristics of this distribution: the canonical mass peak and its dependence on spectral type, the high-mass tail associated with mergers or with massive crystallized objects, and the low-mass population linked to binary evolution. Mapping the mass distribution as a function of spectral type, temperature, and Galactic component (thin disk, thick disk, halo) will provide key insights into the physical processes that shape the observed white dwarf population. Moreover, the mass distribution is directly connected to the initial-to-final mass relation and the initial mass function, and thus to the mass range of stars that contributed to the local white dwarf population over Galactic history.

\subsection{The Luminosity Function}

The white dwarf luminosity function  has been recognized for decades as a powerful tool in the analysis of white dwarf populations, providing insights into Galactic evolution, age, and star formation history \citep[e.g.][and references therein]{garcia-berro2016}. To develop its full potential, however, the  white dwarf luminosity function must be constructed from unbiased and nearly complete samples with accurately determined luminosities. This, in turn, requires precise spectroscopic measurements of effective temperature, surface gravity, and atmospheric composition, which are needed to interpolate within reliable cooling models and thus obtain accurate luminosities.

A spectroscopically calibrated white dwarf luminosity function will improve constraints on the onset and termination of star formation episodes, the age of the local Galactic disk, and the contributions of mergers and exotic evolutionary channels at the faint end. It will also provide an essential benchmark for testing models of crystallization, phase separation, and energy release in cooling white dwarfs.  Moreover, as recently highlighted by \citet{isern2022}, the  white dwarf luminosity function also serves as a unique laboratory for fundamental physics, enabling constraints on additional cooling mechanisms, exotic particles, or possible variations in physical constants.

\subsection{The Magnetic Field Distribution}

Magnetic white dwarfs represent a significant component of the population--about 10-20\%--, and exhibit magnetic field strengths spanning from $\sim$kG to $1\,$GG \citep{ferrario2015,bagnulo2022}. The distribution of these field strengths--and their correlation with atmospheric composition, mass, and kinematics-- will shed light on the still-debated origin of magnetism in compact stars.  A statistically significant magnetic sample is therefore crucial to discriminate among the proposed formation channels, including fossil fields inherited from the progenitor, dynamo amplification during late stellar evolution, and magnetic-field generation in binary mergers \citep{bagnulo2022,camisassa2024b,moss2025}.

\section{Spectroscopic Needs and ESO Capabilities}

\subsection{Spectroscopic Requirements for Gaia White Dwarfs}

Despite the transformative impact of {\em Gaia} on white dwarf astrophysics, the very low spectral resolution of the mission prevents a reliable determination of spectral types. {\em Gaia} spectra are primarily useful for distinguishing DAs from non-DAs, and in some cases identifying a few broad spectral subclasses, but they do not provide the level of detail required for robust atmospheric classification. Accurate spectroscopic characterization is therefore essential for inferring the atmospheric composition and physical properties that govern white dwarf evolution.

High-quality spectra spanning the near-UV, optical, and near-infrared are required for determining atmospheric composition, distinguishing H-rich, He-rich, mixed, detecting carbon-enriched atmospheres, or metal-polluted envelopes,  and for resolving the line profiles required to obtain precise measurements of effective temperature, surface gravity, and magnetic field strength.

Current and ongoing multi-fiber, large area surveys on 4-m telescopes like DESI \citep{cooper2023}, 4MOST \citep{dejong2019}, and WEAVE \citep{jin2024} will contribute to the spectroscopic confirmation of most white dwarfs down to a {\em Gaia} magnitude of $G \approx 20$\,mag, thus achieving a full-sky statistically significant low-resolution ($R \approx 5000$) sample of white dwarfs. Several existing instruments are available to the single-object follow-up of -- mostly --  the brightest white dwarfs. Near-UV and optical coverage is provided to the ESO community by FORS2, X-shooter, UVES, and ESPRESSO, offering resolutions from low to ultra-high for detailed atmospheric diagnostics, as well as (spectro-) polarimetry for accurate magnetic-field characterization. Looking ahead, the ELT instruments HARMONI, HIRES, and METIS will deliver unprecedented optical to near-infrared sensitivity and resolution, allowing comprehensive spectroscopic characterization of the most interesting white dwarfs across the {\em Gaia} HR diagram.

\subsection{The need for multi-object survey spectrographs}

In the coming years, we expect a large increase of white dwarf samples in the southern hemisphere thanks to the 10-year long  Legacy Survey for Space and Time (LSST) at the Vera C. Rubin Observatory \citep{zekljko2020}. This six-band, time-resolved photometric survey will scan the entire observable sky every few nights, allowing to determine parallaxes and proper-motions for Galactic objects. The proper-motion selection from the long-baseline measurements is expected to allow the identification of a few million white dwarfs down to the 25$^{\rm th}$ magnitude in the $r$ band, including a few 10\,000s halo members \citep{fantin2020}. Thus, the magnitude depth of LSST will allow the full identification of a complete white dwarf sample within 100-pc of the Sun, and will allow identifications up to $\approx 3$\,kpc away.

While the existing and planned facilities operating in the 2030s will permit the follow-up of single objects, a full unbiased spectroscopic sample will only be possible with multi-fiber survey instruments on large-aperture telescopes. The existing multi-fiber spectrographs on 4-m telescopes are at most suitable for low-resolution ($R\approx 5000$) spectroscopy of $\approx 20$\,mag white dwarfs, and their high-resolution ($R\approx 10\,000$) capability is limited to 16$^{\rm th}$ magnitude targets. Thus, future instruments will need to operate on telescopes of 10--15\,m mirrors, allowing the low-resolution follow-up of fainter targets down to the magnitude limits of LSST and a magnitude complete high-resolution spectroscopic sample down to $\sim 20$\,mag. Extending the high-resolution samples, in particular, will enable to measure accurate radial velocities of white dwarfs \citep{napiwotzki2020} that are key to measure the space velocities of white dwarfs, thus enabling their kinematic association to Galactic populations.

\section*{Technology and Data Handling Requirements}
\label{sec:tech}

In order to fully characterize a spectroscopic sample of white dwarfs that will be identified by the LSST, in the 2040s we foresee using a massively multiplexing instrument that will inherit the technological and scientific achievements of the existing instruments on 4-m telescopes. Such an instrument will require a larger collecting area, at least 8-m but ideally between 10--15\,m, that will allow to obtain low-resolution ($R = 2000$--5000) spectra for white dwarfs as faint as $G = 23$--25\,mag and high-resolution spectra ($R = 10\,000$--40\,000) for brighter objects of $G \leq 20$\,mag.

Low-resolution spectroscopy is the \textit{cheapest} classification tool for white dwarfs, which allows to measure their atmospheric composition (H, He, C, and other contaminating heavy elements), effective temperature and surface gravity with sufficient precision and accuracy via model fitting \citep{bergeron1992} with state-of-the-art spectral libraries \citep{Koester2010,tremblay2013,cukanovaite2021} down to a signal-to-noise ratio of about $\mathrm{SNR} \sim 20$. Thus, a survey instrument will be beneficial to observe around one million objects in a typical 5-year survey timeline.

A few such multiplexed instruments exist on large 8-m telescopes \citep[e.g., the Prime Focus Spectrograph of the Subaru telescope;][]{tamura2016}, or have been proposed for the use as survey instruments at the end of the 2030s and in the 2040s decade, like the Maunakea Spectroscopic Explorer \citep[MSE;][]{mse2019} and the Wide-field Spectroscopic Telescope \citep[WST;][]{mainieri2024}. These or similar survey instruments will enable the identification of the largest possible samples of white dwarfs, as their low-resolution optical spectral coverage (3600--10\,000\,\AA) is ideal for observing and classifying these objects.

High-resolution, instead, serves more specific tasks such as the identification and the time-resolved follow-up of double-degenerate systems, the measurement of accurate radial velocities, the identification of weak magnetic fields ($10$--$100$\,kG), the precise measurement of small traces of accreted metals, among many other possible uses. Hence, a high-resolution instrument for the follow-up of white dwarfs will ideally cover the H$\alpha$ region, to measure radial-velocity variations from its narrow core in DA white dwarfs or weak magnetic fields. Blue-sensitive instruments ($\lambda \leq 3400$\,\AA), potential successors of CUBES, will also be useful to measure abundances of accreted metals or the presence of small atmospheric traces of C.  Other optical spectral windows could also cover strong He\,{\sc i} lines, or metal lines such as C, and Mg, and Ca that are detected in some white dwarfs dredging up core material or accreting rocky exo-asteroids and planetesimals. The ELT will achieve high precision even for the faintest white dwarfs identified by the LSST, however, a smaller facility with mirror apertures in the range of 10--15\,m, operating as survey instrument, could also support a high-resolution spectrograph that can provide follow-up of interesting white dwarfs.

\begin{acknowledgements}
    This research was partially supported by the AGAUR/Generalitat de Catalunya grant SGR-386/2021 and the Spanish MINECO grant, PID2023-148661NB-I00. RR acknowledges support from Grant RYC2021-030837-I and MEC acknowledges grant RYC2021-032721-I, both funded by MCIN/AEI/ 10.13039/501100011033 and by “European Union NextGeneration EU/PRTR”.
\end{acknowledgements}


\bibliographystyle{aa}

\bibliography{references}

@ARTICLE{althaus2010,
       author = {{Althaus}, Leandro G. and {C{\'o}rsico}, Alejandro H. and {Isern}, Jordi and {Garc{\'\i}a-Berro}, Enrique},
        title = "{Evolutionary and pulsational properties of white dwarf stars}",
      journal = {\aapr},
         year = 2010,
        month = oct,
       volume = {18},
       number = {4},
        pages = {471-566},
          doi = {10.1007/s00159-010-0033-1},
archivePrefix = {arXiv},
       eprint = {1007.2659},
 primaryClass = {astro-ph.SR},
       adsurl = {https://ui.adsabs.harvard.edu/abs/2010A&ARv..18..471A}
}

@ARTICLE{bagnulo2022,
       author = {{Bagnulo}, Stefano and {Landstreet}, John D.},
        title = "{Multiple Channels for the Onset of Magnetism in Isolated White Dwarfs}",
      journal = {\apjl},
         year = 2022,
        month = aug,
       volume = {935},
       number = {1},
          eid = {L12},
        pages = {L12},
          doi = {10.3847/2041-8213/ac84d3},
archivePrefix = {arXiv},
       eprint = {2208.02655},
 primaryClass = {astro-ph.SR},
       adsurl = {https://ui.adsabs.harvard.edu/abs/2022ApJ...935L..12B}
}

@ARTICLE{bauer2020,
       author = {{Bauer}, Evan B. and {Schwab}, Josiah and {Bildsten}, Lars and {Cheng}, Sihao},
        title = "{Multi-gigayear White Dwarf Cooling Delays from Clustering-enhanced Gravitational Sedimentation}",
      journal = {\apj},
         year = 2020,
        month = oct,
       volume = {902},
       number = {2},
          eid = {93},
        pages = {93},
          doi = {10.3847/1538-4357/abb5a5},
archivePrefix = {arXiv},
       eprint = {2009.04025},
 primaryClass = {astro-ph.SR},
       adsurl = {https://ui.adsabs.harvard.edu/abs/2020ApJ...902...93B}
}

@ARTICLE{bedard2024,
       author = {{B{\'e}dard}, Antoine},
        title = "{The spectral evolution of white dwarfs: where do we stand?}",
      journal = {\apss},
         year = 2024,
        month = apr,
       volume = {369},
       number = {4},
          eid = {43},
        pages = {43},
          doi = {10.1007/s10509-024-04307-5},
archivePrefix = {arXiv},
       eprint = {2405.01268},
 primaryClass = {astro-ph.SR},
       adsurl = {https://ui.adsabs.harvard.edu/abs/2024Ap&SS.369...43B}
}

@ARTICLE{bedard2024b,
       author = {{B{\'e}dard}, Antoine and {Blouin}, Simon and {Cheng}, Sihao},
        title = "{Buoyant crystals halt the cooling of white dwarf stars}",
      journal = {\nat},
         year = 2024,
        month = mar,
       volume = {627},
       number = {8003},
        pages = {286-288},
          doi = {10.1038/s41586-024-07102-y},
archivePrefix = {arXiv},
       eprint = {2409.04419},
 primaryClass = {astro-ph.SR},
       adsurl = {https://ui.adsabs.harvard.edu/abs/2024Natur.627..286B}
}

@ARTICLE{bergeron1992,
       author = {{Bergeron}, P. and {Saffer}, Rex A. and {Liebert}, James},
        title = "{A Spectroscopic Determination of the Mass Distribution of DA White Dwarfs}",
      journal = {\apj},
         year = 1992,
        month = jul,
       volume = {394},
        pages = {228},
          doi = {10.1086/171575},
       adsurl = {https://ui.adsabs.harvard.edu/abs/1992ApJ...394..228B}
}

@ARTICLE{bergeron2019,
       author = {{Bergeron}, P. and {Dufour}, P. and {Fontaine}, G. and {Coutu}, S. and {Blouin}, S. and {Genest-Beaulieu}, C. and {B{\'e}dard}, A. and {Rolland}, B.},
        title = "{On the Measurement of Fundamental Parameters of White Dwarfs in the Gaia Era}",
      journal = {\apj},
         year = 2019,
        month = may,
       volume = {876},
       number = {1},
          eid = {67},
        pages = {67},
          doi = {10.3847/1538-4357/ab153a},
archivePrefix = {arXiv},
       eprint = {1904.02022},
 primaryClass = {astro-ph.SR},
       adsurl = {https://ui.adsabs.harvard.edu/abs/2019ApJ...876...67B}
}

@ARTICLE{bergeron2022,
       author = {{Bergeron}, P. and {Kilic}, Mukremin and {Blouin}, Simon and {B{\'e}dard}, A. and {Leggett}, S.~K. and {Brown}, Warren R.},
        title = "{On the Nature of Ultracool White Dwarfs: Not so Cool after All}",
      journal = {\apj},
         year = 2022,
        month = jul,
       volume = {934},
       number = {1},
          eid = {36},
        pages = {36},
          doi = {10.3847/1538-4357/ac76c7},
archivePrefix = {arXiv},
       eprint = {2206.03174},
 primaryClass = {astro-ph.SR},
       adsurl = {https://ui.adsabs.harvard.edu/abs/2022ApJ...934...36B}
}

@ARTICLE{blouin2021,
       author = {{Blouin}, Simon and {Daligault}, J{\'e}r{\^o}me and {Saumon}, Didier},
        title = "{$^{22}$Ne Phase Separation as a Solution to the Ultramassive White Dwarf Cooling Anomaly}",
      journal = {\apjl},
         year = 2021,
        month = apr,
       volume = {911},
       number = {1},
          eid = {L5},
        pages = {L5},
          doi = {10.3847/2041-8213/abf14b},
archivePrefix = {arXiv},
       eprint = {2103.12892},
 primaryClass = {astro-ph.SR},
       adsurl = {https://ui.adsabs.harvard.edu/abs/2021ApJ...911L...5B}
}

@ARTICLE{camisassa2016,
       author = {{Camisassa}, Mar{\'\i}a E. and {Althaus}, Leandro G. and {C{\'o}rsico}, Alejandro H. and {Vinyoles}, N{\'u}ria and {Serenelli}, Aldo M. and {Isern}, Jordi and {Miller Bertolami}, Marcelo M. and {Garc{\'\i}a{\textendash}Berro}, Enrique},
        title = "{The Effect of $^{22}$NE Diffusion in the Evolution and Pulsational Properties of White Dwarfs with Solar Metallicity Progenitors}",
      journal = {\apj},
         year = 2016,
        month = jun,
       volume = {823},
       number = {2},
          eid = {158},
        pages = {158},
          doi = {10.3847/0004-637X/823/2/158},
archivePrefix = {arXiv},
       eprint = {1604.01744},
 primaryClass = {astro-ph.SR},
       adsurl = {https://ui.adsabs.harvard.edu/abs/2016ApJ...823..158C}
}

@ARTICLE{camisassa2017,
       author = {{Camisassa}, Mar{\'\i}a E. and {Althaus}, Leandro G. and {Rohrmann}, Ren{\'e} D. and {Garc{\'\i}a-Berro}, Enrique and {Torres}, Santiago and {C{\'o}rsico}, Alejandro H. and {Wachlin}, Felipe C.},
        title = "{Updated Evolutionary Sequences for Hydrogen-deficient White Dwarfs}",
      journal = {\apj},
         year = 2017,
        month = apr,
       volume = {839},
       number = {1},
          eid = {11},
        pages = {11},
          doi = {10.3847/1538-4357/aa6797},
archivePrefix = {arXiv},
       eprint = {1703.05340},
 primaryClass = {astro-ph.SR},
       adsurl = {https://ui.adsabs.harvard.edu/abs/2017ApJ...839...11C}
}

@ARTICLE{camisassa2021,
       author = {{Camisassa}, Mar{\'\i}a E. and {Althaus}, Leandro G. and {Torres}, Santiago and {C{\'o}rsico}, Alejandro H. and {Rebassa-Mansergas}, Alberto and {Tremblay}, Pier-Emmanuel and {Cheng}, Sihao and {Raddi}, Roberto},
        title = "{Forever young white dwarfs: When stellar ageing stops}",
      journal = {\aap},
         year = 2021,
        month = may,
       volume = {649},
          eid = {L7},
        pages = {L7},
          doi = {10.1051/0004-6361/202140720},
archivePrefix = {arXiv},
       eprint = {2008.03028},
 primaryClass = {astro-ph.SR},
       adsurl = {https://ui.adsabs.harvard.edu/abs/2021A&A...649L...7C}
}

@ARTICLE{Camisassa2023,
       author = {{Camisassa}, Maria and {Torres}, Santiago and {Hollands}, Mark and {Koester}, Detlev and {Raddi}, Roberto and {Althaus}, Leandro G. and {Rebassa-Mansergas}, Alberto},
        title = "{A hidden population of white dwarfs with atmospheric carbon traces in the Gaia bifurcation}",
      journal = {\aap},
         year = 2023,
        month = jun,
       volume = {674},
          eid = {A213},
        pages = {A213},
          doi = {10.1051/0004-6361/202346628},
archivePrefix = {arXiv},
       eprint = {2305.02110},
 primaryClass = {astro-ph.SR},
       adsurl = {https://ui.adsabs.harvard.edu/abs/2023A&A...674A.213C}
}

@ARTICLE{camisassa2024,
       author = {{Camisassa}, Maria and {Baiko}, Denis A. and {Torres}, Santiago and {Rebassa-Mansergas}, Alberto},
        title = "{Exsolution process in white dwarf stars}",
      journal = {\aap},
         year = 2024,
        month = mar,
       volume = {683},
          eid = {A101},
        pages = {A101},
          doi = {10.1051/0004-6361/202348344},
archivePrefix = {arXiv},
       eprint = {2312.00604},
 primaryClass = {astro-ph.SR},
       adsurl = {https://ui.adsabs.harvard.edu/abs/2024A&A...683A.101C}
}

@ARTICLE{camisassa2024b,
       author = {{Camisassa}, M. and {Fuentes}, J.~R. and {Schreiber}, M.~R. and {Rebassa-Mansergas}, A. and {Torres}, S. and {Raddi}, R. and {Dominguez}, I.},
        title = "{Main sequence dynamo magnetic fields emerging in the white dwarf phase}",
      journal = {\aap},
         year = 2024,
        month = nov,
       volume = {691},
          eid = {L21},
        pages = {L21},
          doi = {10.1051/0004-6361/202452539},
archivePrefix = {arXiv},
       eprint = {2411.02296},
 primaryClass = {astro-ph.SR},
       adsurl = {https://ui.adsabs.harvard.edu/abs/2024A&A...691L..21C}
}

@ARTICLE{cummings2018,
       author = {{Cummings}, Jeffrey D. and {Kalirai}, Jason S. and {Tremblay}, P.-E. and {Ramirez-Ruiz}, Enrico and {Choi}, Jieun},
        title = "{The White Dwarf Initial-Final Mass Relation for Progenitor Stars from 0.85 to 7.5 M $_{{\ensuremath{\odot}}}$}",
      journal = {\apj},
         year = 2018,
        month = oct,
       volume = {866},
       number = {1},
          eid = {21},
        pages = {21},
          doi = {10.3847/1538-4357/aadfd6},
archivePrefix = {arXiv},
       eprint = {1809.01673},
 primaryClass = {astro-ph.SR},
       adsurl = {https://ui.adsabs.harvard.edu/abs/2018ApJ...866...21C}
}

@ARTICLE{cheng2019,
       author = {{Cheng}, Sihao and {Cummings}, Jeffrey D. and {M{\'e}nard}, Brice},
        title = "{A Cooling Anomaly of High-mass White Dwarfs}",
      journal = {\apj},
         year = 2019,
        month = dec,
       volume = {886},
       number = {2},
          eid = {100},
        pages = {100},
          doi = {10.3847/1538-4357/ab4989},
archivePrefix = {arXiv},
       eprint = {1905.12710},
 primaryClass = {astro-ph.SR},
       adsurl = {https://ui.adsabs.harvard.edu/abs/2019ApJ...886..100C}
}

@ARTICLE{cooper2023,
       author = {{Cooper}, Andrew P. and {Koposov}, Sergey E. and {Allende Prieto}, Carlos and {Manser}, Christopher J. and {Kizhuprakkat}, Namitha and {Myers}, Adam D. and {Dey}, Arjun and {G{\"a}nsicke}, Boris T. and {Li}, Ting S. and {Rockosi}, Constance and {Valluri}, Monica and {Najita}, Joan and {Deason}, Alis and {Raichoor}, Anand and {Wang}, M.-Y. and {Ting}, Y.-S. and {Kim}, Bokyoung and {Carrillo}, Andreia and {Wang}, Wenting and {Beraldo e Silva}, Leandro and {Han}, Jiwon Jesse and {Ding}, Jiani and {S{\'a}nchez-Conde}, Miguel and {Aguilar}, Jessica N. and {Ahlen}, Steven and {Bailey}, Stephen and {Belokurov}, Vasily and {Brooks}, David and {Cunha}, Katia and {Dawson}, Kyle and {de la Macorra}, Axel and {Doel}, Peter and {Eisenstein}, Daniel J. and {Fagrelius}, Parker and {Fanning}, Kevin and {Font-Ribera}, Andreu and {Forero-Romero}, Jaime E. and {Gazta{\~n}aga}, Enrique and {Gontcho a Gontcho}, Satya and {Guy}, Julien and {Honscheid}, Klaus and {Kehoe}, Robert and {Kisner}, Theodore and {Kremin}, Anthony and {Landriau}, Martin and {Levi}, Michael E. and {Martini}, Paul and {Meisner}, Aaron M. and {Miquel}, Ramon and {Moustakas}, John and {Nie}, Jundan J.~D. and {Palanque-Delabrouille}, Nathalie and {Percival}, Will J. and {Poppett}, Claire and {Prada}, Francisco and {Rehemtulla}, Nabeel and {Schlafly}, Edward and {Schlegel}, David and {Schubnell}, Michael and {Sharples}, Ray M. and {Tarl{\'e}}, Gregory and {Wechsler}, Risa H. and {Weinberg}, David H. and {Zhou}, Zhimin and {Zou}, Hu},
        title = "{Overview of the DESI Milky Way Survey}",
      journal = {\apj},
         year = 2023,
        month = apr,
       volume = {947},
       number = {1},
          eid = {37},
        pages = {37},
          doi = {10.3847/1538-4357/acb3c0},
archivePrefix = {arXiv},
       eprint = {2208.08514},
 primaryClass = {astro-ph.GA},
       adsurl = {https://ui.adsabs.harvard.edu/abs/2023ApJ...947...37C}
}

@ARTICLE{cukanovaite2021,
       author = {{Cukanovaite}, Elena and {Tremblay}, Pier-Emmanuel and {Bergeron}, Pierre and {Freytag}, Bernd and {Ludwig}, Hans-G{\"u}nter and {Steffen}, Matthias},
        title = "{3D spectroscopic analysis of helium-line white dwarfs}",
      journal = {\mnras},
         year = 2021,
        month = mar,
       volume = {501},
       number = {4},
        pages = {5274-5293},
          doi = {10.1093/mnras/staa3684},
archivePrefix = {arXiv},
       eprint = {2011.12693},
 primaryClass = {astro-ph.SR},
       adsurl = {https://ui.adsabs.harvard.edu/abs/2021MNRAS.501.5274C}
}

@ARTICLE{fantin2020,
       author = {{Fantin}, Nicholas J. and {C{\^o}t{\'e}}, Patrick and {McConnachie}, Alan W.},
        title = "{White Dwarfs in the Era of the LSST and Its Synergies with Space-based Missions}",
      journal = {\apj},
         year = 2020,
        month = sep,
       volume = {900},
       number = {2},
          eid = {139},
        pages = {139},
          doi = {10.3847/1538-4357/aba270},
archivePrefix = {arXiv},
       eprint = {2007.01312},
 primaryClass = {astro-ph.GA},
       adsurl = {https://ui.adsabs.harvard.edu/abs/2020ApJ...900..139F}
}

@ARTICLE{ferrario2015,
       author = {{Ferrario}, Lilia and {de Martino}, Domitilla and {G{\"a}nsicke}, Boris T.},
        title = "{Magnetic White Dwarfs}",
      journal = {\ssr},
         year = 2015,
        month = oct,
       volume = {191},
       number = {1-4},
        pages = {111-169},
          doi = {10.1007/s11214-015-0152-0},
archivePrefix = {arXiv},
       eprint = {1504.08072},
 primaryClass = {astro-ph.SR},
       adsurl = {https://ui.adsabs.harvard.edu/abs/2015SSRv..191..111F}
}

@ARTICLE{Koester2010,
       author = {{Koester}, D.},
        title = "{White dwarf spectra and atmosphere models}",
      journal = {\memsai},
         year = 2010,
        month = jan,
       volume = {81},
        pages = {921-931},
       adsurl = {https://ui.adsabs.harvard.edu/abs/2010MmSAI..81..921K}
}

@ARTICLE{dejong2019,
       author = {{de Jong}, R.~S. and {Agertz}, O. and {Berbel}, A.~A. and {Aird}, J. and {Alexander}, D.~A. and {Amarsi}, A. and {Anders}, F. and {Andrae}, R. and {Ansarinejad}, B. and {Ansorge}, W. and {Antilogus}, P. and {Anwand-Heerwart}, H. and {Arentsen}, A. and {Arnadottir}, A. and {Asplund}, M. and {Auger}, M. and {Azais}, N. and {Baade}, D. and {Baker}, G. and {Baker}, S. and {Balbinot}, E. and {Baldry}, I.~K. and {Banerji}, M. and {Barden}, S. and {Barklem}, P. and {Barth{\'e}l{\'e}my-Mazot}, E. and {Battistini}, C. and {Bauer}, S. and {Bell}, C.~P.~M. and {Bellido-Tirado}, O. and {Bellstedt}, S. and {Belokurov}, V. and {Bensby}, T. and {Bergemann}, M. and {Bestenlehner}, J.~M. and {Bielby}, R. and {Bilicki}, M. and {Blake}, C. and {Bland-Hawthorn}, J. and {Boeche}, C. and {Boland}, W. and {Boller}, T. and {Bongard}, S. and {Bongiorno}, A. and {Bonifacio}, P. and {Boudon}, D. and {Brooks}, D. and {Brown}, M.~J.~I. and {Brown}, R. and {Br{\"u}ggen}, M. and {Brynnel}, J. and {Brzeski}, J. and {Buchert}, T. and {Buschkamp}, P. and {Caffau}, E. and {Caillier}, P. and {Carrick}, J. and {Casagrande}, L. and {Case}, S. and {Casey}, A. and {Cesarini}, I. and {Cescutti}, G. and {Chapuis}, D. and {Chiappini}, C. and {Childress}, M. and {Christlieb}, N. and {Church}, R. and {Cioni}, M.-R.~L. and {Cluver}, M. and {Colless}, M. and {Collett}, T. and {Comparat}, J. and {Cooper}, A. and {Couch}, W. and {Courbin}, F. and {Croom}, S. and {Croton}, D. and {Daguis{\'e}}, E. and {Dalton}, G. and {Davies}, L.~J.~M. and {Davis}, T. and {de Laverny}, P. and {Deason}, A. and {Dionies}, F. and {Disseau}, K. and {Doel}, P. and {D{\"o}scher}, D. and {Driver}, S.~P. and {Dwelly}, T. and {Eckert}, D. and {Edge}, A. and {Edvardsson}, B. and {Youssoufi}, D.~E. and {Elhaddad}, A. and {Enke}, H. and {Erfanianfar}, G. and {Farrell}, T. and {Fechner}, T. and {Feiz}, C. and {Feltzing}, S. and {Ferreras}, I. and {Feuerstein}, D. and {Feuillet}, D. and {Finoguenov}, A. and {Ford}, D. and {Fotopoulou}, S. and {Fouesneau}, M. and {Frenk}, C. and {Frey}, S. and {Gaessler}, W. and {Geier}, S. and {Gentile Fusillo}, N. and {Gerhard}, O. and {Giannantonio}, T. and {Giannone}, D. and {Gibson}, B. and {Gillingham}, P. and {Gonz{\'a}lez-Fern{\'a}ndez}, C. and {Gonzalez-Solares}, E. and {Gottloeber}, S. and {Gould}, A. and {Grebel}, E.~K. and {Gueguen}, A. and {Guiglion}, G. and {Haehnelt}, M. and {Hahn}, T. and {Hansen}, C.~J. and {Hartman}, H. and {Hauptner}, K. and {Hawkins}, K. and {Haynes}, D. and {Haynes}, R. and {Heiter}, U. and {Helmi}, A. and {Aguayo}, C.~H. and {Hewett}, P. and {Hinton}, S. and {Hobbs}, D. and {Hoenig}, S. and {Hofman}, D. and {Hook}, I. and {Hopgood}, J. and {Hopkins}, A. and {Hourihane}, A. and {Howes}, L. and {Howlett}, C. and {Huet}, T. and {Irwin}, M. and {Iwert}, O. and {Jablonka}, P. and {Jahn}, T. and {Jahnke}, K. and {Jarno}, A. and {Jin}, S. and {Jofre}, P. and {Johl}, D. and {Jones}, D. and {J{\"o}nsson}, H. and {Jordan}, C. and {Karovicova}, I. and {Khalatyan}, A. and {Kelz}, A. and {Kennicutt}, R. and {King}, D. and {Kitaura}, F. and {Klar}, J. and {Klauser}, U. and {Kneib}, J.-P. and {Koch}, A. and {Koposov}, S. and {Kordopatis}, G. and {Korn}, A. and {Kosmalski}, J. and {Kotak}, R. and {Kovalev}, M. and {Kreckel}, K. and {Kripak}, Y. and {Krumpe}, M. and {Kuijken}, K. and {Kunder}, A. and {Kushniruk}, I. and {Lam}, M.~I. and {Lamer}, G. and {Laurent}, F. and {Lawrence}, J. and {Lehmitz}, M. and {Lemasle}, B. and {Lewis}, J. and {Li}, B. and {Lidman}, C. and {Lind}, K. and {Liske}, J. and {Lizon}, J.-L. and {Loveday}, J. and {Ludwig}, H.-G. and {McDermid}, R.~M. and {Maguire}, K. and {Mainieri}, V. and {Mali}, S. and {Mandel}, H.},
        title = "{4MOST: Project overview and information for the First Call for Proposals}",
      journal = {The Messenger},
         year = 2019,
        month = mar,
       volume = {175},
        pages = {3-11},
          doi = {10.18727/0722-6691/5117},
archivePrefix = {arXiv},
       eprint = {1903.02464},
 primaryClass = {astro-ph.IM},
       adsurl = {https://ui.adsabs.harvard.edu/abs/2019Msngr.175....3D}
}

@ARTICLE{doherty2017,
       author = {{Doherty}, Carolyn L. and {Gil-Pons}, Pilar and {Siess}, Lionel and {Lattanzio}, John C.},
        title = "{Super-AGB Stars and their Role as Electron Capture Supernova Progenitors}",
      journal = {\pasa},
         year = 2017,
        month = nov,
       volume = {34},
          eid = {e056},
        pages = {e056},
          doi = {10.1017/pasa.2017.52},
archivePrefix = {arXiv},
       eprint = {1703.06895},
 primaryClass = {astro-ph.SR},
       adsurl = {https://ui.adsabs.harvard.edu/abs/2017PASA...34...56D}
}

@ARTICLE{garcia-berro2016,
       author = {{Garc{\'\i}a-Berro}, Enrique and {Oswalt}, Terry D.},
        title = "{The white dwarf luminosity function}",
      journal = {\nar},
         year = 2016,
        month = jun,
       volume = {72},
        pages = {1-22},
          doi = {10.1016/j.newar.2016.08.001},
archivePrefix = {arXiv},
       eprint = {1608.02631},
 primaryClass = {astro-ph.SR},
       adsurl = {https://ui.adsabs.harvard.edu/abs/2016NewAR..72....1G}
}

@ARTICLE{garcia-zamora2025,
       author = {{Garc{\'\i}a-Zamora}, Enrique Miguel and {Torres}, Santiago and {Rebassa-Mansergas}, Alberto and {Ferrer-Burjachs}, Aina},
        title = "{A random forest spectral classification of the Gaia 500 pc white dwarf population}",
      journal = {\aap},
         year = 2025,
        month = jun,
       volume = {699},
          eid = {A3},
        pages = {A3},
          doi = {10.1051/0004-6361/202554414},
archivePrefix = {arXiv},
       eprint = {2505.05560},
 primaryClass = {astro-ph.SR},
       adsurl = {https://ui.adsabs.harvard.edu/abs/2025A&A...699A...3G}
}

@ARTICLE{gentilefusillo2021,
       author = {{Gentile Fusillo}, N.~P. and {Tremblay}, P.-E. and {Cukanovaite}, E. and {Vorontseva}, A. and {Lallement}, R. and {Hollands}, M. and {G{\"a}nsicke}, B.~T. and {Burdge}, K.~B. and {McCleery}, J. and {Jordan}, S.},
        title = "{A catalogue of white dwarfs in Gaia EDR3}",
      journal = {\mnras},
         year = 2021,
        month = dec,
       volume = {508},
       number = {3},
        pages = {3877-3896},
          doi = {10.1093/mnras/stab2672},
archivePrefix = {arXiv},
       eprint = {2106.07669},
 primaryClass = {astro-ph.SR},
       adsurl = {https://ui.adsabs.harvard.edu/abs/2021MNRAS.508.3877G}
}

@ARTICLE{isern2022,
       author = {{Isern}, J. and {Torres}, S. and {Rebassa-Mansergas}, A.},
        title = "{White dwarfs as Physics laboratories: lights and shadows}",
      journal = {Frontiers in Astronomy and Space Sciences},
         year = 2022,
        month = jan,
       volume = {9},
          eid = {6},
        pages = {6},
          doi = {10.3389/fspas.2022.815517},
archivePrefix = {arXiv},
       eprint = {2202.02052},
 primaryClass = {astro-ph.HE},
       adsurl = {https://ui.adsabs.harvard.edu/abs/2022FrASS...9....6I}
}

@ARTICLE{jimenezesteban2023,
       author = {{Jim{\'e}nez-Esteban}, F.~M. and {Torres}, S. and {Rebassa-Mansergas}, A. and {Cruz}, P. and {Murillo-Ojeda}, R. and {Solano}, E. and {Rodrigo}, C. and {Camisassa}, M.~E.},
        title = "{Spectral classification of the 100 pc white dwarf population from Gaia-DR3 and the virtual observatory}",
      journal = {\mnras},
         year = 2023,
        month = feb,
       volume = {518},
       number = {4},
        pages = {5106-5122},
          doi = {10.1093/mnras/stac3382},
archivePrefix = {arXiv},
       eprint = {2211.08852},
 primaryClass = {astro-ph.SR},
       adsurl = {https://ui.adsabs.harvard.edu/abs/2023MNRAS.518.5106J}
}

@ARTICLE{mccook1999,
       author = {{McCook}, George P. and {Sion}, Edward M.},
        title = "{A Catalog of Spectroscopically Identified White Dwarfs}",
      journal = {\apjs},
         year = 1999,
        month = mar,
       volume = {121},
       number = {1},
        pages = {1-130},
          doi = {10.1086/313186},
       adsurl = {https://ui.adsabs.harvard.edu/abs/1999ApJS..121....1M}
}

\end{document}